**Title**: Estimating the construct validity of Principal Components Analysis

**Authors**: Thomas M.H. Hope [1,2], Cathy J. Price [1], Ajay Halai [3], Carola Salvi [2], Jenny Crinion [4], Merel Keijsers [2], Christoph Sperber [5], Howard Bowman [6]

[1] Department of Imaging Neuroscience, Institute of Neurology, University College London, 12 Queen Square, London, WC1N 3AR, United Kingdom

[2] Department of Psychology and Social Sciences, John Cabot University, 00165 Rome, Italy

[3] MRC Cognition and Brain Sciences Unit, University of Cambridge, Cambridge CB2 7EF, UK

[4] Institute of Cognitive Science, Department of Experimental Psychology, University College London, London, WC1N 3AR, UK

[4] Universitätsklinik für Neurologie Inselspital, Freiburgstr. 16, 3010 Bern, Switzerland

[6] School of Psychology, University of Birmingham, Birmingham B15 2TT, UK

**Corresponding author**: Thomas Hope, t.hope@ucl.ac.uk

# Abstract

In many scientific disciplines, the features of interest cannot be observed directly, so must instead be inferred from observed behaviour. Latent variable analyses are increasingly employed to systematise these inferences, and Principal Components Analysis (PCA) is perhaps the simplest and most popular of these methods. Here, we examine how the assumptions that we are prepared to entertain, about the latent variable system, mediate the likelihood that PCA-derived components will capture the true sources of variance underlying data. As expected, we find that this likelihood is excellent in the best case, and robust to empirically reasonable levels of measurement noise, but best-case performance is also: (a) not robust to violations of the method's more prominent assumptions, of linearity and orthogonality; and also (b) requires that other subtler assumptions be made, such as that the latent variables should have varying importance, and that weights relating latent variables to observed data have zero mean. Neither variance explained, nor replication in independent samples, could reliably predict which (if any) PCA-derived components will capture true sources of variance in data. We conclude by describing a procedure to fit these inferences more directly to empirical data, and use it to find that components derived via PCA from two different empirical neuropsychological datasets, are less likely to have meaningful referents in the brain than we hoped.

# 1. Introduction

In many scientific disciplines, the features of interest cannot be observed directly, so must instead be inferred from observed behaviour. In the study of the brain, for example, those 'features of interest' might be the function of dissociable cognitive sub-systems, and the 'observed behaviour' might be accuracies and / or reaction times recorded in behavioural tasks that are thought to employ these subsystems. In this and many other fields, researchers commonly use latent variable analyses to systematise the inverse inference from observed data to features of interest[1-9]. Principal Components Analysis (PCA) is one of the simplest and most popular of these methods. Here, we examine the likelihood that this inverse inference will succeed: i.e., that PCA-derived components will capture the true sources of variance underlying observed data. As expected, we find that this likelihood – what we call the construct validity of PCA-derived components – is excellent in the best case. But this excellence is more sensitive to subtler assumptions, about the latent variable system, than we thought – including assumptions that might be counter-intuitive in empirical practice. We also find that natural proxies for construct validity, including the variance that PCA-derived components explain, and the likelihood that they will replicate in independent samples, are unreliable. Finally, we propose a procedure to make more targeted inferences about empirical PCA-derived components – and use it to show that their construct validity might be lower than expected.

PCA operates by iteratively deriving variables / components that capture as much of the variance of the original data as possible. Each PCA-derived component is derived from data after regressing out variance captured by previous components. The result is an ordered set of uncorrelated (orthogonal) eigenvectors or components of the original data. The variance that each successive component explains, of the original data, is smaller than that explained by the last. Later PCA-derived components might explain very little variance, and perhaps be dominated by noise [10]. This distinction, between more and less 'interesting' components, is what makes PCA so powerful as a tool for data compression and (potentially) de-noising [11-13].

However, in practice, PCA is not just used for data compression and de-noising. In the field of population genetics, for example, PCA is now a standard feature of analyses aiming to draw substantive conclusions about populations (e.g., concerning ancestry and kinship [14,15]). PCA-derived components are offered for download from publicly accessible datasets such as the UK BioBank [16], built into analysis pipelines for functional brain imaging data [17-19], and used to understand the molecular dynamics of materials such as graphene [20,21]. Increasingly,

PCA is also used to interpret profiles of behavioural impairment, consequent to brain damage, in terms of dysfunction at the level of separable cognitive sub-systems or functions [5,6,22-24]. Applications like this make little sense if PCA-derived components are interpreted only as convenient, compressed representations of slices of the original data's variance. These applications also presume that those slices are meaningful: i.e., that they capture something important or illuminating about the data-generating (latent variable) system. More formally, they presume that PCA will have reasonable construct validity in practice, in the sense that PCA-derived components will be reasonably likely to capture the true sources of variance underlying observed, multivariate data. Indeed, the use of factor rotation, a common post-processing step for PCA, is often expressly motivated by the aim of deriving components that are more naturally interpretable as capturing real latent variables. Here, we examine the construct validity of PCA, in detail.

PCA assumes that observed data are linear mixtures of orthogonal latent variables; this much is already well-known, as is the risk that data departing from these assumptions might confound the method[8]. But quite what this means in practice, is often hard to judge. For example, though non-linearity is evidently fundamental to brain function[25,26], linear analyses of brain data still often appear to be useful[27-30]. This might be because functional brain dynamics are really more linear than expected[31], or because linear methods can often approximate non-linear signals reasonably well[32], or both. In other words, linear methods often appear to work well, even when we expect that the associations under study might not be linear. Similar logic can be applied to the assumption that latent variables are orthogonal. Minor violations of this assumption might just not matter much in practice. And if latent variables are so strongly correlated that our analyses cannot distinguish them, perhaps that distinction does not matter very much, either?

Logic like this allows us to convince ourselves that PCA might get close enough to be useful, even when we are not confident that the latent system respects its assumptions. And any residual doubt might be allayed by replication, if the same or similar components are derived from independent samples[33,34].

Here, we attempt to go beyond this comforting but vague intuition, that PCA can often get close enough to be useful, to ask exactly how its construct validity varies as we vary the (prominent and less prominent) assumptions we make about the latent variable system. We find that, while PCA's best-case construct validity is excellent, it is also: (a) not particularly robust to violations of its prominent assumptions; (b) more sensitive than expected to subtler assumptions about the latent system (which may be counter-intuitive in empirical practice); and (c) potentially very low, even when PCA-derived components explain much of the original variance, or have been replicated in independent samples. Finally, we estimate the construct validity of latent variables derived via PCA from two empirical datasets – and find that those estimates are low.

## 2. Methods

We present the results of a series of analyses applying PCA to synthetic data, derived via known functions from known latent variables. The first is an omnibus analysis that systematically permutes all of our simulation hyperparameters – variables that define a space of reasonable architectures for data-generating mechanisms. This analysis aims to explore how variations in those hyperparameters mediate the construct validity of the (PCA-derived) results. In the second analysis, we search for hyperparameters that yield data whose covariance structure matches that of empirical datasets as closely as possible, and thereby estimate the construct validity of latent variables derived from those empirical data, via PCA.

Figure 1 displays a schematic illustration of the kind of data-generating mechanisms that our simulation hyperparameters will define. In each simulation we: (a) define latent variables according to the simulation hyperparameters, (b) derive the specified number of observed variables from those latent variables, (c) analyse the observed variables with PCA, and (d) compare the derived components to the real latent variables that were actually used to produce the observed data. That last step – the comparison between derived components and real latent variables – is where we measure construct validity.

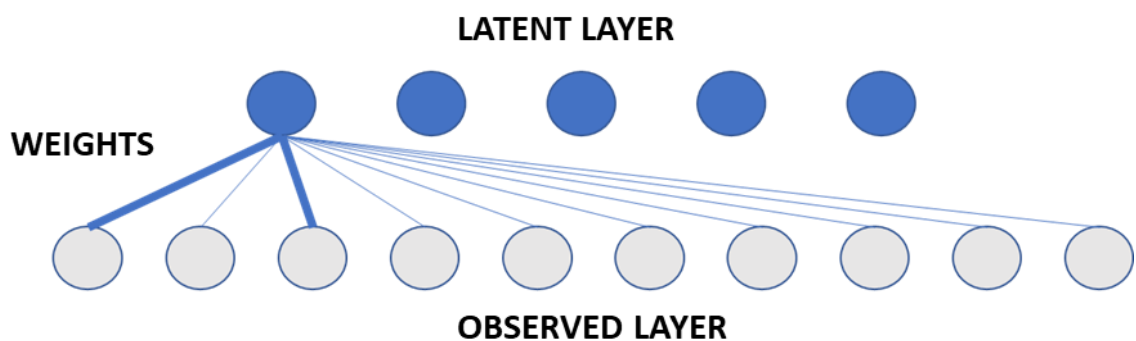

**Figure 1:** A schematic illustration of the structure used to generate synthetic observed data. Every unit in the latent layer is connected to every unit in the observed layer, via a weighted

connection. Given a pre-defined weight matrix, the observed values for a given participant are calculated by propagating latent variable values through those weights to the observed layer (some simulation hyperparameter configurations then add a sigmoid transform), and then adding any measurement noise.

### 2.1. Measuring construct validity

We define construct validity as the probability that PCA-derived components can sensibly be said to 'significantly and selectively correlate with' any real latent variable. To pass this test, a PCA-derived component must: (a) be significantly correlated with a real latent variable of the system; (b) be significantly more strongly correlated with that real latent variable, than any other PCA-derived component; and (c) be significantly more strongly correlated with that real latent variable, than with any other real latent variable.

Criteria (b-c) require comparisons between dependent correlation coefficients: i.e., correlations between each of two different variables, and a common third variable. We use Wilcox's bootstrap procedure to make these comparisons[35]. Specifically, we calculate the difference between the (absolute) correlation coefficients for each of 500 bootstrap samples of the original variables. A significant difference is observed if 1-$\alpha$% of those differences are greater than zero. The focus on absolute correlation coefficients is appropriate because the directions of PCA-derived components are arbitrary. To make the process more efficient, we only compare the strongest and second-strongest correlations for each row and column of the (PCA-derived components x real latent variables) correlation coefficient matrix. The dependent correlations comparisons were run as 1-tailed tests, where $\alpha = 5\%$.

This analysis is restricted to those PCA-derived components whose eigenvalues are >1: i.e., surpassing the Kaiser-Guttman threshold that is popularly used to distinguish 'more important' from 'less important' principal components [36]. For each PCA-derived component that surpasses the threshold, we tests whether it significantly and selectively correlates with some real latent variable of the system. The probability of that correspondence – what we refer to as the *per-component construct validity* – is the proportion of simulations with the same hyperparameters, where each component passes this test. We measure construct validity both at the level of individual, PCA-derived components, and as an average of the per-component percentages. These summary figures are required because every simulation involves at least an element of randomness. Therefore, there will be variance in the construct validity measures of PCA-derived components across different simulations with the same hyperparameters.

Notably, our measures of construct validity might interact with the number of components that we derive. To see why, consider the situation in which we derive 3 principal components from data derived from only 2 real latent variables. In this case, only 2 components can possibly pass our significant-and-selective-correlation test, so average per-component construct validity can never reach 100%. We control for these effects by considering at most as many PCA-derived components than there are real latent variables in the system under study. Another control is the use of the relatively strict Kaiser-Guttmann criterion to count PCA-derived components. In fact, this criterion could encourage the derivation of fewer components than real latent variables[37,38], which might conceivably inflate our measures of construct validity. So, we also attempt to mitigate that tendency by employing at least 10 times as many behavioural scores as the maximum number of real latent variables that we consider. This strategy was found was effective in the recent past[38].

### 2.2. The omnibus analysis

Our omnibus analysis is designed to survey how per-component construct validity varies over a space of reasonable hyperparameters for data-generating systems. The hyperparameters include:

(a) The sample size (100 or 1,000): i.e., sample sizes that might be considered either large or 'very large' in many psychological and neuropsychological studies;

(b) The number of latent variables (2, 5, or 10): i.e., allowing for both low-dimensional and higher-dimensional systems;

(c) The number of observed variables (100): i.e., 10 times as many observed variables as the maximum number of latent variables considered;

(d) The distribution of latent variable values (uniform or Gaussian random);

(e) The distribution of latent-to-observed variable weights (uniform or Gaussian random);

(f) The scale of latent variable values (zero mean or positive only);

(g) The scale of weight values (zero mean or positive only);

(h) Measurement noise on the observed variables, with standard deviation equal to 0, or 1 times the standard deviation of the observed variables without noise, The maximum value for this noise will encourage average worst-case test retest reliability (or correlation between the same variables observed at different times) of ~0.7. This is a reasonable minimum level of reliability for task scores employed in neuropsychology, the field from which our two empirical datasets are drawn. (e.g., [39]).

(i) Correlations between the latent variables (none, or average r = 0.5);

(j) The importance of the latent variables: equally important or with monotonically decreasing importance. For an ordered set of latent variables, we model monotonically decreasing importance by dividing all of the weights from the $i$-th latent variable by $i$. The result is that latent variable importance follows an exponential decay function, which is somewhat similar to the shape of scree plots often observed in empirical studies using PCA.

(k) Whether the latent-to-observed variable relationship is linear or non-linear (sigmoidal); and

(l) Whether rotation was used (we consider varimax and promax rotation, as these are common in the literature: e.g., [4,9]).

Each hyperparameter combination specifies a procedure for generating latent variable values, and for deriving synthetic observed data from those latent variables. We analyse the latter with PCA, and compare the resulting, PCA-derived components to the real latent variables using the test described previously. And as mentioned previously, since every hyperparameter combination includes a degree of randomness, we conduct 1,000 simulations per combination, and report summary results. We fix the number of observed variables to 100 because this is ten times the largest number of latent variables considered. As mentioned in the last section (2.1), we recently found that this ratio encourages accurate latent space dimensionality estimation, so should minimise any confounding effects of dimensionality estimation errors on our measures of construct validity[38].

### 2.3. Inverse inference from empirical data

In this analysis, we aim to estimate the construct validity of components derived from particular, empirical datasets. We do this by inverse inference from the empirical data itself. Specifically, we search for simulation hyperparameters that produce synthetic observed data, whose covariance matches that of empirical observed data, i.e., we minimise the Mean Absolute Error (MAE) between the two. The focus on covariance is appropriate because this is what drives PCA. We then estimate the construct validity of empirical PCA-derived components from the resultant, best-fitting models.

We run the search in two ways: first as a grid search, and then via Bayesian optimization. The grid search considers the same hyperparameter configurations as considered in the omnibus analysis. However, we do expand that hyperparamter space a little, to include all real system dimensionalities in the range 1-10. This is because it seemed important to allow that estimates of empirical data dimensionality might be correct, and neither of our empirical datasets yielded estimates of 2, 5, or 10 components, as we considered in the omnibus analysis.

Leveraging the greater efficiency of the Bayesian search, we further expand the search space to include: (a) any real latent dimensionality in the range 1-50; (b) all real numbered average latent variable correlations in the range 0-1 (instead of just 0 or 0.5); and (c) measurement noise of all standard deviations in the range 0-1 (instead of just 0 and 1). We ran the search for 300 iterations, minimising the MAE between the covariance matrices of synthetic, observed data, and the covariance matrix of empirical, observed data.

Finally, in all analyses aiming to fit simulation hyperparameters to empirical data, the sample size and number of observed scores were always fixed to values defined by the empirical data.

### 2.4. Empirical data

We illustrate the empirical inverse inference procedure for two neuropsychological datasets, which record profiles of behavioural impairment consequent to stroke. Stroke is a common, abrupt brain injury that occurs when blood vessels in the brain are either blocked (ischaemic stroke) or burst, leading to bleeding (haemorrhagic stroke). In either case, normal blood flow is interrupted, depriving downstream neurons of oxygen. This can lead to cell death, disrupting any cognitive functions whose premorbid implementation employed / depended on those cells. To capture the potentially enormous variety of consequent impairments, stroke researchers typically use wide-ranging batteries of behavioural tasks, designed so that different tasks depend on different combinations of putative, cognitive functions (e.g., [39,40]).

Profiles of behavioural impairment, recorded from these task batteries, are then used to infer what cognitive impairment(s) each participant might have suffered. For example, a patient who cannot read single words might have failed to (A) understand the task instructions, to (B) see or (C) recognise the word, to (D) associate the word with its sound, or to (E) create or (F) execute a plan to articulate that sound. If the same patient can repeat the same words when they hear them, we can (tentatively) locate their cognitive impairment to some combination of functions B, C, and / or D. And still more targeted inferences could be made if we consider the same patient's performance on more tasks. This is the kind of inverse inference that we mentioned in the introduction, which is increasingly being approached using PCA.

The first of our two datasets refers to patients drawn from the Predicting Language Outcomes and Recovery (PLORAS) dataset [41]. This dataset includes task scores from participants whose language and cognitive skills were assessed at University College London (UK), at a wide range of times post-stroke (from <3 months to >10 years). The second dataset was acquired much sooner after stroke (< 2 weeks), by researchers at the University of Washington in St. Louis [5]. Both datasets are cross-sectional, and include scores from wide-ranging batteries of standardised, behavioural tasks; here, we consider only those scores that refer to language impairments (aphasia) after stroke.

# 3. Results

## 1.1. Omnibus analysis

### 1.1.1. Hyper-parameter main effects

Table 1 reports how variation in our simulations' hyper-parameters affects the mean per-component reconstruction probabilities across all simulations – including the results of Kruskall-Wallis tests for the significance of that variation.

| Factor | Chi squared | p | Factor Level | | |
|---|---|---|---|---|---|
| | | | 1 | 2 | 3 |
| **No. of latent variables (2, 5, 10)** | 1508.43 | <0.001 | 57.9% | 34.9% | 19.8% |
| **Latent variable orthogonality (r ~ 0, r ~ 0.5)** | 422.91 | <0.001 | 45.4% | 29.6% | |
| **Sample size (100, 1000)** | 416.83 | <0.001 | 29.9% | 45.1% | |
| **Latent-to-observed linearity (Linear, Sigmoid)** | 168.77 | <0.001 | 43.1% | 31.9% | |
| **Latent variable importance (Equal, Monotonic)** | 90.91 | <0.001 | 33.4% | 41.6% | |
| **Positive only weights (No, Yes)** | 50.95 | <0.001 | 40.3% | 34.7% | |
| **Latent variable distribution (Uniform, Gaussian)** | 17.04 | <0.001 | 39.1% | 36.0% | |
| **Weights distribution (Uniform, Gaussian)** | 1.18 | 0.278 | 37.2% | 37.9% | |
| **Use factor rotation (No, Varimax, Promax)** | 0.61 | 0.737 | 39.1% | 36.6% | 36.9% |
| **Noise standard deviation (0, or 1)** | 0.05 | 0.819 | 37.6% | 37.4% | |

| | | | | |
|---|---|---|---|---|
| **Positive only latent variables (No, Yes)** | 0.01 | 0.921 | 37.6% | 37.5% |

Table 1: Non-parametric main effects of hyperparameters on the construct validity of PCA-derived components, and the average per-component construct validity for each level of each factor. Effects are ordered by effect size (largest first). The variable values for each factor are indicated in brackets next to the name of that factor.

Average per-component construct validity is *higher* in simulations with: (a) fewer real latent variables; which are (b) uncorrelated; with (c) larger sample sizes; related to observed data via (d) zero-mean weights; and (e) with variable average influence on observed data; which are (f) linear mixtures of those (g) randomly uniformly distributed latent variables. Average construct validity was not significantly affected by the use of factor rotation, the presence of measurement noise, the scale of latent variable values, or the distribution of latent-to-observed variable weights.

Since factor rotation is often used to derive more interpretable components, it might be surprising that the procedure did not significantly improve the construct validity of PCA-derived components in our omnibus analysis. In fact, there was evidence that factor rotation can help, as expected. For example, promax rotation caters to situations in which we expect the latent variables to be correlated [9]. In this case, average construct validity was better when promax rotation was used (34.5%) than when no rotation was used (20.6%). This lower-level effect is masked in Table 1, because best-case construct validity occurs principally when the latent variables are uncorrelated, and no factor rotation is employed.

### 1.1.2. Best-case construct validity

As expected, given the method's long history and popularity, the best-case construct validity of PCA is excellent. Simulations with perfect (100%) average construct validity were observed for latent systems with 2 or 5 real latent variables only, though best-case performance for simulations with 10 real latent variables was still very high, at 97.2%. Interestingly, in low-dimensional systems (with just 2 real latent variables), perfect construct validity was possible even when the latent variables were correlated, and / or the latent-to-observed data relationship was non-linear: i.e., violating PCA's most prominent assumptions.

All of best-case simulations included: (a) zero-mean latent-to-observed variable weights, and (b) latent variables with unequal average influences on observed data. None of the best-case simulations employed factor rotation. Best-case performance for 10-latent-variable systems required these same features, and also that those two more prominent assumptions of PCA (of linearity and orthogonality) should be respected.

### 1.1.3. Relaxing the best-case assumptions

Across all real latent-system dimensionalities that we considered, best-case construct validity occurred when observed data were: (a) linear mixtures; of (b) orthogonal latent variables; with (c) unequal latent variable importance; and (d) zero-mean latent-to-observed weights. Here, we explore how construct validity falls when these assumptions are relaxed. Table 2 reports this next-best-case performance: i.e., the best observed average construct validity when each of those best-case assumptions is relaxed (singly or in pairs). None of these second-best-case simulations employed factor rotation.

| Best-case constraint(s) relaxed | No. of Latent Variables | | |
|---|---|---|---|
| | 2 | 5 | 10 |
| **None (i.e., best-case)** | 100.0% | 100.0% | 97.2% |
| **Linearity (to nonlinear)** | 100.0% | 80.7% | 75.4% |
| **Orthogonality (to correlated)** | 100.0% | 77.1% | 55.8% |
| **Latent variable importance (to equal)** | 97.9% | 79.0% | 60.0% |
| **Weight scale (to positive only)** | 80.4% | 77.1% | 75.4% |
| **Non-linear AND correlated latent variables** | 100.0% | 72.9% | 55.8% |
| **Non-linear AND equally import latent variables** | 97.9% | 75.3% | 55.9% |
| **Non-linear latent variables AND positive-only weights** | 80.4% | 72.9% | 75.4% |
| **Correlated, equally important latent variables** | 79.0% | 77.1% | 55.8% |
| **Correlated latent variables and positive only weights** | 78.9% | 77.1% | 55.8% |
| **Equally important latent variables and positive only weights** | 79.1% | 76.2% | 59.9% |

Table 2: 'Next-best-case' construct validity, or the maximum (average per-component) construct validity observed when the specified best-case assumptions are relaxed.

In 2-dimensional systems, PCA had good construct validity even when those systems were non-linear, or included correlated latent variables, or latent variables of equal importance. However, more significant costs were observed in higher-dimensional systems when any of these three best-case assumptions were changed. The effect of changing the fourth best-case assumption, that latent-to-observed variable weights should have zero mean (i.e., instead ensuring that the weights were all positive-only), was worse for 2-dimensional systems, than when the other assumptions were relaxed, but this effect did not appear to worsen much further for 5- and 10-dimensional systems.

### 1.1.4. Variance explained does not predict construct validity

It would be convenient if the variance that a PCA-derived component explained, of the original data, was a reliable proxy for the construct validity of that component. Construct validity is significantly correlated with variance explained across all simulations that we considered ($r = 0.09$, $p<0.001$), but unfortunately this summary analysis is confounded because lower-dimensional latent variable systems yielded fewer PCA-derived components that both explained more variance, and had higher construct validity. When we perform the analysis separately for latent systems with different dimensionalities, the hoped-for relationship is significant and positive for 2-dimensional systems ($r = 0.24$, $p <0.001$), marginally negative for 5-dimensional systems ($r = -0.03$, $p = 0.001$), and negative (i.e., lower variance explained predicting higher construct validity) for 10-dimensional systems ($r = -0.13$, $p < 0.001$).

Going beyond these correlations, we can imagine actually trying to distinguish 'probably valid' PCA-derived components in practice. Perhaps the most natural way to do this would be with a threshold; components that explain supra-threshold variance could be considered valid, while those below it might be dismissed as invalid. Unfortunately, there simply is no threshold, in our results, which could be effective in this sense. We illustrate the point in Figure 2, which plots the probability that components will have at least reasonably good construct validity, for 4 definitions of what 'at least reasonably good' means. This chance never rises above 35%, even if we are prepared to accept a permissive threshold of 60%, for acceptable construct validity.

Notably, after initially increasing to around 20% variance explained, the lines in figure 2 all then decrease. The implication is that the chances of good construct validity actually fall with higher thresholds for variance explained. This is because those higher thresholds increasingly select for only the first PCA-derived component, and this component often has lower construct validity than the second (in 41.3% of our simulations, or 1905 / 4608). In 69 simulations – all involving positive-only latent-to-observed variable weights – the construct validity of the first principal component was <10%, even when the construct validity of the second component was >90%.

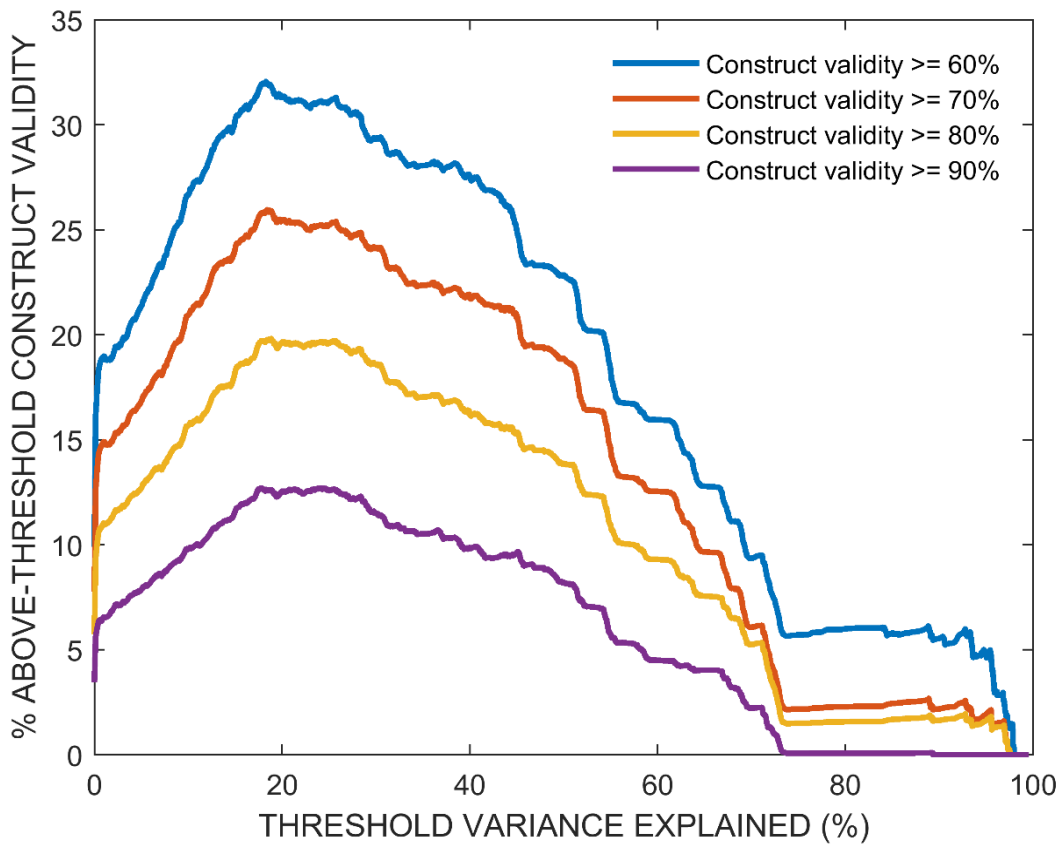

Figure 2: The proportion (y-axis) of PCA-derived components that explain at least as much variance as a specified threshold (x-axis), and which also display construct validity above some threshold. 4 lines are displayed, corresponding to 4 different construct validity thresholds (60%, 70%, 80%, and 90%).

### 1.1.5. Replication does not predict construct validity

In search of an alternative – and hopefully more reliable – proxy for construct validity, we now turn to 'replication'. This is the intuition that PCA-derived components might be more likely to capture true sources of variance if the same or similar components are also derived from further, independent samples. The same intuition, that statistical stability might be a proxy for construct validity, can motivate significance tests for PCA [42,43]. To test the intuition, we re-ran the omnibus analysis again, drawing not one but two synthetic datasets from each data-generating mechanism. We performed PCA on both datasets, and then correlated each PCA-derived component with each of the original scores. The result is two sets of (component x observed) coefficient matrices. A PCA-derived component from the first dataset is considered to be 'replicated' if its coefficients are 'significantly and selectively' correlated with the coefficients of a component derived from the second dataset: i.e., we are using the same logic to measure replication as we used to measure construct validity.

Figure 3 plots the relationship between: (a) replication rates; and (b) the probability that construct validity will be 'good enough', for 4 different definitions of that threshold. Even with a permissive threshold of 60% for acceptable construct validity, no replication rate guarantees that we will surpass that threshold even 40% of the time. This chance falls to less than 20%, for any replication rate, if we want to ensure our PCA-derived components have at least 80% construct validity. And perhaps surprisingly, PCA-derived components that replicate perfectly (100% of the time) are actually less likely to meet any threshold for minimum construct validity, than components with less perfect replication.

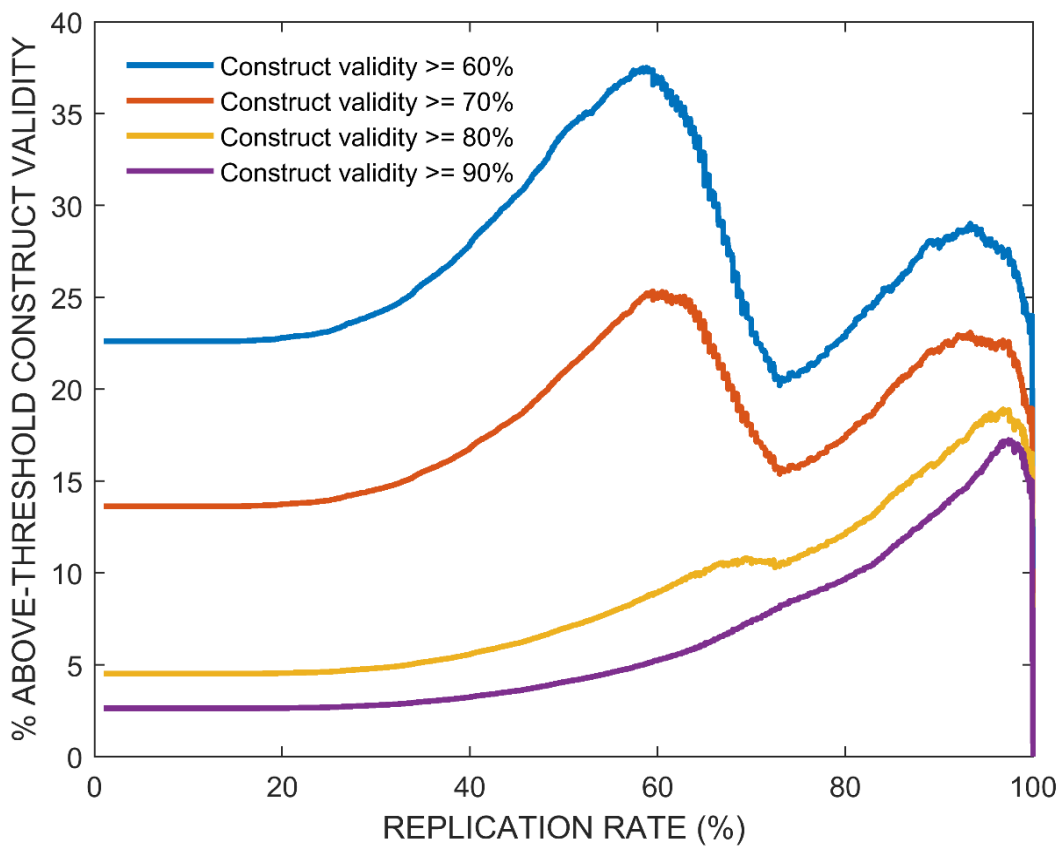

Figure 3: The proportion (y-axis) of PCA-derived components that replicate at least often as a specified threshold (x-axis), which also display construct validity above some threshold. 4 lines are displayed, corresponding to 4 different construct validity thresholds: 60%, 70%, 80%, and 90%.

### 1.1.6. Summary of the omnibus analysis results

The preceding sections illustrate that, while potentially very high, the construct validity of PCA-derived components can vary, substantially, depending on the details of the mechanism that was used to generate observed data. Naturally, construct validity suffers when the method's more prominent assumptions – linearity and orthogonality – are violated. But even when they are respected, there were many simulation configurations, corresponding to assumptions we might reasonably make of an empirical system under study, where construct validity was also low. Finally, neither high variance explained, nor high replication rates, could be used to reliably predict whether PCA-derived components might actually correspond to real latent variables.

However, this rather abstract space of possibility does not directly address what will be the most important question for many, which is how to estimate the construct validity of their own, empirical, PCA-derived components.

### 1.2. Inverse inference from empirical data

Here, we estimate the construct validity of empirical PCA-derived components. We do this by searching for simulation hyperparameters that generate synthetic observed data whose covariance matches that of empirical data as closely as possible. Best-fitting simulation hyperparameters then drive the estimates of construct validity. We illustrate that inference for two empirical datasets from the same discipline: stroke research. In each case, the observed data are scores assigned on the basis of patients' performance in a range of tasks designed to assess the severity of post-stroke impairment. And in each case, we are estimating the probability that components derived from these behavioural data, via PCA, will capture true sources of variance underlying those behavioural data (i.e., their construct validity).

Dataset 1 includes 29 language scores from 1,367 participants. 3 latent variables were derived from this dataset, using the Kaiser-Guttman criterion. Dataset 2 includes 9 language scores from 122 participants, yielding just one principal component that surpasses the Kaiser-Guttman criterion [36]. Table 3 displays the best-fitting simulation hyperparameters for these empirical data, after searching the space using: (a) a grid search on the hyperparameters considered for the omnibus analysis (expanded to consider all latent space dimensionalities in the range 1-10, as described in the Methods); and (b) a Bayesian optimisation procedure, which operates on expanded hyperparameter ranges as indicated in the table. Estimated construct validity for the empirical PCA-derived components is also included: three per-component estimates for Dataset 1, and one estimate for Dataset 2. Each estimate is conducted three times: without factor rotation, with varimax rotation, and with promax rotation.

| Factor | Dataset 1 | | Dataset 2 | |
| --- | --- | --- | --- | --- |
| | Grid search | Bayesian Optimisation | Grid search | Bayesian Optimisation |
| **No. of participants** | 1367 | 1367 | 122 | 122 |
| **No. of observed scores** | 29 | 29 | 9 | 9 |
| **No. of latent variables** | 10 | 48 | 10 | 48 |
| **Latent variable mean correlation** | 0 | <0.001 | 0 | 0.001 |
| **Latent variable importance** | All equal | All equal | All equal | All equal |
| **Weights scale** | Positive only | Positive only | Positive only | Positive only |
| **Latent variable scale** | Positive only | Zero mean | Positive only | Zero mean |
| **Weights distribution** | Random uniform | Random uniform | Random uniform | Random uniform |
| **Latent variable distribution** | Gaussian | Gaussian | Gaussian | Gaussian |
| **Latent-to-observed linearity** | Sigmoid | Sigmoid | Linear | Linear |
| **Noise standard deviation** | 0 | 0.01 | 0 | 0.47 |
| **Covariance error (MAE)** | 0.127 | 0.126 | 0.0919 | 0.068 |
| **Construct validity (%)** | 1.7, 53.0, 52.8 | 0.0, 5.8, 6.7 | 1.7 | 0.3 |
| **Construct validity + varimax rotation** | 19.4, 29.5, 26.1 | 0, 0.8, 0.8 | 13.3 | 0.0 |
| **Construct validity + promax rotation** | 15.2, 21.0, 24.2 | 0.9, 1.6, 0.6 | 11.8 | 0.8 |

Table 3: Best-fitting simulation hyperparameters for two empirical datasets, identified either via a grid search of the space employed for the omnibus analysis (768 combinations), or via a Bayesian optimisation of a wider range of those hyperparameters (300 iterations). In both cases, we measure goodness of fit as the Mean Absolute Error between: (a) the covariance matrix for the empirical data; and (b) the covariance matrices of the synthetic data. Remembering that there are 1,000 covariance matrices per hyperparameter configuration, this comparison yields 1,000 MAE measures per configuration, and the table reports the average of those 1,000 measures. The rows below report the per-component construct validity estimates (3 percentages for Dataset 1, and 1 percentage for Dataset 2) for each simulation configuration.

The best-fitting simulation for Dataset 1, as identified via grid search, yielded per-component construct validity estimates of 1.7%, 53.0%, and 52.8% for each of the three components that we derived from that dataset. Factor rotation improved the construct validity of the first PCA-derived component (though it was still less than 20% at best), but at the cost of lower construct validity for the other two components. Notably, this pattern of generally low construct validity was consistent across the 100 best-fitting simulations, where only 16.2% (146) of the construct validity estimates were >50%, and the average estimates were: 14.5%, 40.9%, and 42.9%, respectively. The best-fitting simulation identified via Bayesian search also yielded low estimates of construct validity. And the same pattern of results was also evident for Dataset 2: see Table 3. In this case, just 4.7% of the construct validity estimates for the best-fitting 100 models was >50%.

## 2. Discussion

One important strength of PCA is that it *just works*. Largely irrespective of the data we have, or the assumptions we make of those data, PCA will usually generate simple results, quickly. Our results here suggest that this facility might also be the method's most important weakness – at least if we aim to interpret PCA-derived components.

None of our analyses and results are new. We have merely applied principles that have long been discussed by researchers in other disciplines, who study the problem of the 'identifiability' of latent systems (e.g., [44-48]). In some fields, the limitations of PCA are perhaps already so well known that they no longer need to be discussed[49,50]. Notwithstanding its best-case construct validity, for some (perhaps many), PCA is just so obviously unsuited to inferences about latent variable structure, that the results presented here might seem redundant. But in the psychological and neuropsychological studies that our results here most naturally address, we suggest that those limits still need to be rediscovered – as they have been, recently, in other fields such as genetics[51] imaging neuroscience[52].

As expected, best-case construct validity for PCA-derived components, was excellent. Best-case performance was also robust to reasonable variations in measurement noise magnitude – or at least to variations that we consider to be reasonable in the study of psychology and neurology. However, best-case performance depended on assumptions beyond the better-known requirements for linearity and orthogonality: namely, unequal latent variable importance and zero-mean latent-to-observed variable weights. When any of these 4 assumptions were violated, excellent construct validity could only be maintained if the real system was low-dimensional – and only then, if most were respected. In practice, we probably cannot measure latent system dimensionality directly or estimate it reliably[38]. So in practice, there is always a risk that our empirical system might be higher-dimensional than we think. In turn, this implies that there is always a risk that our PCA-derived components might have lower construct validity than expected. Applying our inverse inference method to two empirical neuropsychological datasets, we estimated that components derived from those datasets have low construct validity. And replication is apparently no defence against these doubts.

This disconnection, between construct validity and replication, is curious. Any feature of a system that persists in replication, should intuitively tell us something meaningful about that system. If we can use PCA-derived components to predict post-stroke outcomes for new stroke patients, for example [22], it is natural to conclude that these components must capture something illuminating or meaningful about those patients. And yet, apparently, this conclusion could often be wrong. This is because PCA is a deterministic method, which will therefore identify the same or similar latent variables in different datasets with the same or similar covariance. In this sense, PCA can be predictive if data covariance is stable, entirely irrespective of whether PCA-derived components have meaningful referents in the latent variable system.

The key limitation of this work is that it is contingent on the models (or data-generating mechanisms) that we have considered. We cannot be sure that we have considered the right models for any specific, empirical domain, and our conclusions might change if we considered different models. For example, the percentages in Table 1 are aggregates across the space of models that we considered. We designed that space to include models that both: (a) allowed for best-case construct validity; and (b) seemed at least potentially reasonable for a variety of different empirical domains. But for any specific domain, this model-space will likely include (or over-represent) models that make little or no empirical sense. In the study of post-stroke outcomes, for example, it is counter-intuitive to include models without measurement noise (i.e., measurement noise standard deviation = 0). This is because behavioural measures of post-stroke impairment are known to be imperfectly reliable (e.g., [39,40]). In this sense, neither of our grid searches yielded empirically plausible best-fitting models (Table 3). That said, this concern is mitigated by the fact that increasing measurement noise had no significant effect on construct validity (Table 1), and because our Bayesian searches considered models with more plausible measurement noise magnitudes. Nevertheless, there might well be models which generate still better fit to the empirical data than those that we discovered here, which are also better suited to identification via PCA.

Moreover, the inference that drives those empirical analyses – from observed data covariance, through simulation hyperparameters, to construct validity – is ill-posed. With enough degrees of freedom, many different data-generating mechanisms, driving different construct validity estimates, might fit observed data equally well. In other words, our current estimates of the construct validity of empirical PCA-derived components are best guesses, given current methods and data, which could yet be revised.

Nevertheless, analyses like ours do at least draw explicit connections between the assumptions that we are prepared to entertain, about latent variable systems, and the likelihood that PCA-derived components might capture real latent variables. These analyses can illustrate previously unknown (or less well-known) constraints, such as the requirements for varying latent variable importance, and zero-mean latent-to-observed variable weights, to support best-case construct validity. In retrospect, the variability requirement seems obvious, because data points that are linear mixtures of 2 equally scaled random latent variables should describe a circular point-cloud, from which the directionality of the original latent variables is clearly undefined. The same logic equally applies in higher dimensions, when the latent variables are independently and identically distributed. The requirement for zero-mean weights was more surprising to us, and may be to others too, but is also apparently well-known in some fields (Park et al, 2023).

Judgements about how likely these requirements are to be reflected in empirical data, will differ depending on the empirical system under study. In stroke outcomes research, we suggest that all of the best-case assumptions might be suspect. First, there is every reason to believe that cognitive systems' effects on behavioural performance are non-linear – and that this non-linearity is more complex than the non-linearity that our models employ [25,26]. Second, post-stroke cognitive impairments are likely often correlated because stroke induced lesions are constrained by the cerebrovascular tree, so that a bleed or blockage of any vessel will interrupt downstream blood flow (e.g., often damaging clusters of neighbouring brain regions [37,53]). Third, dramatically varying latent variable importance, in stroke, is arguably precisely what the design of post-stroke task batteries is designed to avoid: i.e., by systematically varying the engagement of putative cognitive functions across behavioural tasks. That said, equal latent importance is probably equally suspect in this domain. Reasonable variations in this feature of our models (which will presumably interact with sample size) might be a fruitful, future expansion of this work. And fourth, zero-mean weights imply that roughly half of those weights will be negative, which in turn implies that better or more preserved cognitive function will act to reduce or impair behavioural performance roughly half of the time. This implication is counter-intuitive, at least in patients with neurological disorders.

Finally, if the latent system is 'cognition', as it might be in stroke outcomes research, then we should probably allow that it could be higher-dimensional than any system considered in the simulations presented here. Though hardly conclusive, it is noteworthy that Bayesian-optimisation-based best-fit hyperparameters for both of our empirical datasets specified high-dimensional latent systems. These results might have been different if we had applied a complexity penalty to the objective function for that search. Lower-dimensional systems encourage higher construct validity, so adding such a penalty might yield over-optimistic estimates of construct validity. Nevertheless, this might be a fruitful direction for future research.

Our results discourage the assumption that PCA will get close enough to be useful, at least in the study of post-stroke cognitive impairment, and perhaps in many other fields as well. One obvious alternative is to employ different latent variable analyses methods. For example, Independent Components Analysis has long been a popular alternative to PCA in studies where latent variables are expected to be correlated [54,55]. And autoencoder-based analyses have proved to be useful when data are non-linear functions of latent variables [56,57]. The use of alternative methods does not release us from the need to think about likely construct validity, but it should expand the range of assumptions under which good construct validity can be expected.

Our results naturally beg the question of why reported results, derived via PCA, often seem to be so sensible? If PCA-derived components probably have such low construct validity, why do so many researchers find that they are consistent with expectations? Some of us recently reported a result like this: a component that loads on verbal articulation task scores assigned to stroke patients, and which correlates with damage to motor regions associated with speech using fMRI in the undamaged brain[24]. Perhaps we have been lucky, and the construct validity of empirical PCA-derived components is higher than our results here suggest. Or perhaps this trend is amplified by the file-drawer effect[58], whereby sensible-seeming results are over-represented in the literature, relative to more confusing or inconsistent results. Or perhaps we are collectively just very good at confabulating sensible-seeming interpretations, largely regardless of what variance our PCA-derived components really capture. Or perhaps the truth is somewhere in between these extremes. Our results cannot answer these questions definitively, but we hope that they will encourage work that might bring answers closer, and improve the confidence with which we can interpret latent variable analyses in empirical practice.